\documentclass{amsart}
\usepackage{amssymb}

\usepackage{graphicx}
\usepackage{amscd}

\theoremstyle{plain}

\numberwithin{equation}{section}

\begin{document}
\title[Epidemic thresholds on scale free graphs]{Epidemic thresholds on scale-free graphs: the interplay between exponent and
preferential choice}
\author{Ph.Blanchard}
\address{A.One: Universit\"{a}t Bielefeld, Fakult\"{a}t f\"{u}r Physik, 33619
Bielefeld, Universt\"{a}tsstr.25, Germany---\\
A.~Two: National Center for Theoretical Sciences, Physics Division, NTHU
101.Section2, Hsinchu, Taiwan 300, R.O.C.----\\
A.Three: Universit\"{a}t Bielefeld, Fakult\"{a}t f\"{u}r Mathematik, 33619
Bielefeld, Universit\"{a}tsstr.25, Germany\\
tkrueger@physik.uni-bielefeld.de---}
\email{blanchard@physik.uni-bielefeld.de, chchang@phys.cts.nthu.edu.tw,\linebreak
tkrueger@physik.uni-bielefeld.de}
\author{C.-H.Chang}
\author{T. Kr\"{u}ger}
\date{22th May 2003}

\begin{abstract}
We study epidemic threshold properties in a scale-free random graph model.
We show via a branching process approximation that the divergence of the
second moment of the degree distribution is equivalent to the absence of an
epidemic threshold. We study further the relation between diameter and
epidemic threshold. Absence of an epidemic threshold happens precisely when
a positive fraction of the nodes form a cluster of bounded diameter.
\end{abstract}

\maketitle

\section{Introduction}

In the last decade there was an enormous increase of interest in phenomena
which show for characteristic observables a power-law behavior. Power-law
distributions are often called scale-free distributions due to the absence
of characteristic sizes. Most prominent under the huge variety of examples
became systems which show so-called selforganized criticality (SOC), \cite{1}
\cite{5}. The basic idea to explain the appearance of a
power-law-distribution is via the linkage to phase-transitions where the
appearance of scale-free structures is known since a long time. Despite many
efforts only little is known about the origin of power-law-distributions on
a mathematical base. Partially due to the increasing knowledge of the
structure of the internet there is a growing number of articles studying the
structure of networks with a power law distribution for the degree (see \cite
{2} and the references therein).

In this article we want to discuss a specific model of scale-free-graphs
based on a partner\ choice strategy and its epidemic threshold properties.
Related questions found recently much interest due to the possible absence
of epidemic thresholds for a certain range of power law exponents \cite{6} 
\cite{7}. We were mainly motivated to our studies by the paper of
Pastor-Satorras and Vespignani \cite{7}, where on largely heuristic grounds
the absence of an epidemic threshold for power-law-graphs with exponents
less than 3 was claimed. Since the argumentation in this paper was quite
doubtful from a mathematical point of view but somehow intuitively
convincing we wanted to study the same question on more solid grounds. To
get enough theoretical control we developed a new model of \
power-law-graphs which seems to be of independent interest. Due to the
strong independence properties - as far as this is possible for a scale-free
degree-distribution - we were able to carry out a rigorous analysis of
certain aspects of the model. For a certain range of parameters we get a
power-law distribution for the total degree. For other parameter values
there is an interesting domain where one has a fat tail degree distribution
with gaps. Here the integrated tail distribution is still of power-law form.
In both cases the divergence of the second moment of the degree distribution
is indeed equivalent to the absence of an epidemic threshold. We furthermore
examine the relation between divergence of the second moment of the degree
distribution and diameter-like properties of the relevant graph-spaces.
Namely one has in the case of absence of an asymptotic epidemic threshold
the striking property that there is a huge (meaning a positive fraction of
the whole population) cluster of finite- size independent- diameter. This
clusters carries the epidemics and forms a kind of very small world inside
the small world of the largest connected component which can still be of
logarithmic diameter.

We further study some models with completely different degree distribution
where the absence of an epidemic threshold is not necessary linked to the
divergence of the second moment of the degree-distribution.

\section{ A preferential-choice model of scale-free graphs}

In the following we want to describe the basic structure of the model we
consider in this paper. Our first aim is to construct a family of static
random graph models in which vertex degrees are distributed power-law like,
while edges still have high degree of independence. As usual in random graph
theory we will entirely deal with asymptotic properties in the sense that
the graph size goes to infinity.

We consider graphs with vertex set $V=V_{n}$ $=\left\{ 1,...,n\right\} $
where an edge between the vertices $x$ and $y$ (denoted by $x\sim y$) is
interpreted as a persistent contact between the two nodes. Given $x\in V$,
its degree is denoted by $d(x)$. We will think of edges as generated by a
pair-formation process in which each vertex $x$ - often denoted as an
individual - chooses a set of partners according to a specified $x$%
-dependent rule. Therefore the set of individuals which have contact with a
given vertex $x$ can be divided into two -possible non-disjoint sets: the
set of nodes which are chosen by $x$ himself and the set of nodes which have
chosen $x$ as one of their partners. We call the size of the first set the
outdegree $d_{out}(x)$of $x$ \ and the size of the second one the indegree $%
d_{in}(x)$ of $x$. Obviously $d(x)\leq d_{out}(x)+d_{in}(x)$ and if the
choices are sufficiently independent one can expect equality to hold almost
surely if $\ n\rightarrow \infty $.

We partition the set of vertices $V_{n}$ into groups $\left\{
C_{i}(n)\right\} _{i\geq 1}$ where all members of a group $C_{i}(n)$ choose
exactly $i$ partners by themselves ($d_{out}=i$ on $C_{i}(n)$). Let $%
P_{\alpha }^{1}\left( n,j\right) $ the probability for $x$ to choose a fixed
partner $y\in $ $C_{j}\left( n\right) $ if $n$ partners are available for
the choice and just one choice will be made be 
\begin{equation}
P_{\alpha }^{1}\left( n,j\right) \;=\;A_{\alpha }(n)\mathbb{\ }\frac{%
j^{\alpha }}{n}
\end{equation}
Here $A_{\alpha }(n)$ is a normalization constant such that $A_{\alpha
}(n)\left( \sum_{i\geq 1}\;|C_{i}(n)|\frac{i^{\alpha }}{n}\right) =1$ and $%
\alpha $ is a real parameter. Since we want $A_{\alpha }(n)\rightarrow
A_{\alpha },$as $n\rightarrow \infty $ we need $\sum_{i}|C_{i}(n)|\frac{%
i^{\alpha }}{n}$ to be bounded as a function of $n$ which will impose later
on constraints on the constant $\alpha $. The parameter $\alpha $ acts as an
affinity parameter tuning the tendency to choose a partner with a high
outdegree or low outdegree. If $\alpha =0$, choices are made without any
preferences and $A_{\alpha }(n)\equiv 1$. For $\alpha >0$ the ''highly
active'' individuals are preferred whereas for $\alpha <0$ the ''low
activity'' individuals are favored. From this we obtain the basic
probability 
\begin{equation*}
\Pr [x\text{ from }C_{i}\left( n\right) \text{ has chosen }y\in C_{j}\left(
n\right) \text{ as one of his partners}]\simeq A_{\alpha }(n)\mathbb{\ }%
\frac{i\cdot j^{\alpha }}{n}
\end{equation*}

Concerning the size of the sets $C_{i}(n)$ we will make the following
assumption: 
\begin{equation}
\frac{|C_{i}(n)|}{n}=:p_{i}\left( n\right) \underset{n\rightarrow \infty }{%
\rightarrow }\frac{c_{1}}{i^{\gamma }}.
\end{equation}

With this choice we have to impose the restriction $\alpha <\gamma -1$ to
ensure the convergence of $A_{\alpha }\left( n\right) $. We require
furthermore $\gamma >2$ throughout the paper since otherwise the expected
indegree for individuals from a fixed group would diverge. The basic
probabilities together with the fixed outdegree-distribution define a
probability distribution on each graph with vertex-set $V_{n}$, and
therefore a random graph space $\mathcal{G}_{n}\left( \alpha ,\gamma \right) 
$.

We want to investigate the threshold properties of epidemic processes on
these random graphs as a function of the parameters $\gamma $ and $\alpha $.
Here we are mainly interested in the question whether the epidemics is
asymptotically ($n\rightarrow \infty $) sub-or overcritical. For the
threshold study in this random graph model it is enough to investigate the
epidemic process as a multi-group branching process since typical graphs
have locally a tree structure. Absence of a threshold means here that for
arbitrary small transition probability $\mu >0$ one still gets for $%
n\rightarrow \infty $ an overcritical epidemic process. In section 3 we will
compare the results for the above model with some specific graph models
which have no power-law-distribution but still diverging second moments of
the degree distribution.

First we want to compute the important pairing probabilities which will be
used frequently. We start with the easier case $\alpha =0$. 
\begin{equation}
\Pr \left( x\thicksim y\mid x\in C_{i};y\in C_{k}\right) =\frac{i+k}{n}-%
\frac{ik}{n^{2}}\underset{n\rightarrow \infty }{\longrightarrow }\frac{i+k}{n%
}\text{ \ \ }
\end{equation}

Completely similar one can compute the corresponding probabilities for $%
\alpha \neq 0$. Dropping the simple details we just state the result:

\begin{equation}
\Pr \left( x\thicksim y\mid x\in C_{i}\ ;y\in C_{k}\right) \simeq \frac{%
A_{\alpha }\left( ki^{\alpha }+k^{\alpha }i\right) }{n}\text{ as }%
n\rightarrow \infty .  \label{55}
\end{equation}

It turns out,that the typical graphs in this model still have for $\alpha <2$
a \emph{power-law }distribution for the degree with an exponent which can be
different from the exponent of the outdegree. For $\alpha >2$ we obtain a
degree distribution which follows in mean a power law but has gaps. To
compare both domains we will use the integrated tail distribution $%
F_{k}:=\Pr \left( d\left( x\right) >k\right) .$ We will show that in both
cases we get the same integrated tail distribution. Since we are interested
in the dependence of the epidemic threshold from the power-law exponent of
the total degree distribution we have to analyze how this exponent varies
with the two parameters $\alpha $ and $\gamma .\ $Since the partner choice
is sufficiently random and not too strongly biased toward high degree
individuals (that's the meaning of the condition $\alpha \leq \gamma -1$) it
is easy to see that the indegree distribution of a vertex from group $C_{i}$
converges for $n\rightarrow \infty $ to a Poisson distribution with mean $%
const\cdot i^{\alpha }$. There are essentially two regimes in the parameter
space, one for which the expected indegree is of smaller order than the
outdegree over all groups and one where the indegree is asymptotically of
larger order. In the first case it is clear that the indegree is too small
to have an effect on the degree distribution exponent. In other words: the
set of individuals with degree $k$ consists mainly of individuals whose
outdegree is of order $k$. An easy estimation using the formula for the
pairing probabilities shows that the expected indegree of individuals from
group $i$ is given by $\mathbb{E}\left( d_{in}\left( x\right) \mid x\in
C_{i}\right) \simeq const\cdot i^{\alpha }$ asymptotically. Therefore the
indegree is of smaller order than the outdegree if $\alpha <1$. In the case $%
\gamma -1\geq \alpha \geq 1$ the set of individuals with degree $k$ consists
mainly of individuals from groups with an index of order $k^{\frac{1}{\alpha 
}}$. Furthermore we will show that the condition for absence of an epidemic
threshold is precisely given by the divergence of the second moment of the
degree distribution.

We will state now the main result of the paper which will be shown in the
next sections.

\subsubsection{\protect\bigskip Main Results:}

a) for $\alpha <1$ the random graph $\mathcal{G}_{n}\left( \alpha ,\gamma
\right) $ has a power-law distribution for the degree with exponent $\gamma $

b) for $1\leq \alpha <\min \left( 2;\gamma -1\right) $ the random graph $%
\mathcal{G}_{n}\left( \alpha ,\gamma \right) $ has a power-law distribution
for the degree with exponent $\gamma ^{\prime }=1+\frac{\gamma }{\alpha }-%
\frac{1}{\alpha }$

c) for $2<\alpha <\gamma -1$ the random graph $\mathcal{G}_{n}\left( \alpha
,\gamma \right) $ has a power-law integrated tail distribution with exponent 
$\frac{\gamma }{\alpha }-\frac{1}{\alpha }$

d) there is an epidemic threshold for the parameter set $\{\alpha <\frac{%
\gamma -1}{2},\gamma >3\}$ \ and absence of an asymptotic threshold on the
complement

e) for $\alpha >\frac{\gamma -1}{2}$ or $\gamma <3$ the second moment of the
degree distribution diverges.

\subsection{The degree distribution and second moment estimation}

In this paragraph we want to estimate the degree distribution, respectively
the integrated tail distribution for the parameter domains $1<\alpha <2$ and 
$2<\alpha <\gamma -1$ and give conditions for the divergence of the second
moments. We first note that for $\alpha >1$ the indegree dominates the
outdegree in every group $C_{i}.$ Furthermore for $n\rightarrow \infty $ and
any groupindex $i$ the indegree distribution converges to a Poisson
distribution with expectation $\lambda \left( i\right) =const\cdot i^{\alpha
}.$ Since the variance is of order $i^{\alpha }$ it is clear that for $%
\alpha >2$ and large $i$ there will be essential no overlap between the
indegree distribution in group $i$ and group $i+1.$

We first deal with the indegree distribution for $\alpha <\min \left(
2;\gamma -1\right) $ : 
\begin{eqnarray}
\Pr \left( d_{in}\left( x\right) =k\right) &\simeq &\sum_{i}\Pr \left(
d_{in}\left( x\right) =k\mid x\in C_{i}\right) \cdot \frac{c_{1}}{i^{\gamma }%
} \\
&=&\sum_{i\leq k}\frac{\left( const\cdot i^{\alpha }\right) ^{k}}{k!}%
e^{-const\cdot i^{\alpha }}\cdot \frac{c_{1}}{i^{\gamma }}
\end{eqnarray}
. The essential interval of indices $i$ which contribute for sufficiently
large $k$ to the sum is given by the condition $k\in \left[ const\cdot
i^{\alpha }-a\left( k\right) i^{\frac{\alpha }{2}};const\cdot i^{\alpha
}+a\left( k\right) i^{\frac{\left( 1+\varepsilon \right) \alpha }{2}}\right] 
$ where $a\left( k\right) $ is a slowly growing function in $k$ (for
instance $\log k$ is fine)$.$ Solving for $i$ and denoting the above
constant by $\tilde{A}$ the we get for the boundaries of the essential index
interval $I_{ess}\left( k\right) :=\left[ i_{\min }\left( k\right) ;i_{\max
}\left( k\right) \right] $ : 
\begin{eqnarray}
i_{\min }\left( k\right) &:&=\left( -\frac{a\left( k\right) }{\tilde{A}\cdot
2}+\left( \frac{a^{2}\left( k\right) }{\tilde{A}^{2}\cdot 4}+\frac{k}{\tilde{%
A}}\right) ^{\frac{1}{2}}\right) ^{\frac{2}{\alpha }} \\
i_{\max }\left( k\right) &:&=\left( \frac{a\left( k\right) }{\tilde{A}\cdot 2%
}+\left( \frac{a^{2}\left( k\right) }{\tilde{A}^{2}\cdot 4}+\frac{k}{\tilde{A%
}}\right) ^{\frac{1}{2}}\right) ^{\frac{2}{\alpha }}
\end{eqnarray}
. For the length $\left| I_{ess}\left( k\right) \right| $ of the essential
interval we get therefore: 
\begin{equation}
\left| I_{ess}\left( k\right) \right| =i_{\max }\left( k\right) -i_{\min
}\left( k\right) =\left( \frac{k}{\tilde{A}}\right) ^{\frac{1}{\alpha }%
}\left( \frac{2a\left( k\right) }{\tilde{A}^{\frac{1}{2}}\alpha k^{\frac{1}{2%
}}}+O\left( \frac{1}{k}\right) \right)
\end{equation}
. For $\alpha <2$ it follows that $I_{ess}\left( k\right) \cap I_{ess}\left(
k+1\right) \neq \emptyset $ for all $k$ sufficiently large$.$ In contrast to
this we have for $\alpha >2$ the situation that certain k-values do
essentially not appear since $\left| I_{ess}\left( k\right) \right| $
converges to zero as $k$ tends to infinity and $I_{ess}\left( k\right) $ has
to contain an integer to give a contribution. We estimate now the sum in the
indegree distribution for large $k$ by restricting the summation to the
essential intervals: 
\begin{equation}
\Pr \left( d_{in}\left( x\right) =k\right) \simeq \sum_{i\in I_{ess}\left(
k\right) }\frac{\left( \tilde{A}\cdot i^{\alpha }\right) ^{k}}{k!}e^{-\tilde{%
A}\cdot i^{\alpha }}\cdot \frac{c_{1}}{i^{\gamma }}
\end{equation}
. Using stirlings formula $k!=\left( 2\pi k\right) ^{\frac{1}{2}}\left( 
\frac{k}{e}\right) ^{k}\left( 1+o\left( 1\right) \right) $ and the
approximation $\left( 1+\frac{\varepsilon }{k}\right) ^{k}=e^{\varepsilon -%
\frac{\varepsilon ^{2}}{2k}+o\left( 1\right) }$ for $\left| \varepsilon
\right| \ll k^{\frac{2}{3}}$ we obtain: 
\begin{eqnarray}
\Pr \left( d_{in}\left( x\right) =k\right) &\simeq &\frac{2a\left( k\right) 
}{\alpha }\cdot k^{\frac{1}{\alpha }-\frac{1}{2}}\cdot \frac{1+o\left(
1\right) }{\left( 2\pi k\right) ^{\frac{1}{2}}} \\
&=&\frac{1}{k^{1+\frac{\gamma }{\alpha }-\frac{1}{\alpha }+o\left( 1\right) }%
}
\end{eqnarray}
. We get therefore an asymptotic power law exponent $\gamma ^{\prime }=1+%
\frac{\gamma -1}{\alpha }$ for the indegree distribution which is also the
exponent of the total degree distribution in the case $\alpha >1$ since then 
$\gamma ^{\prime }<\gamma $ and hence the indegree dominates the outdegree.

It is much more easy to estimate the integrated tail distribution. We will
do this again for the indegree distribution and $\alpha >2$ to be able to
compare with the situation for smaller values of $\alpha .$ Clearly one has 
\begin{eqnarray}
F_{k} &=&\Pr \left( d_{in}\left( x\right) >k\right) =\sum_{i}\sum_{x\in
C_{i}}\sum_{l>k}\Pr \left( d_{in}\left( x\right) =l\mid x\in C_{i}\right) \\
&\simeq &\sum_{i}\sum_{l>k}\Pr \left( d_{in}\left( x\right) =l\mid x\in
C_{i}\right) \frac{c_{1}}{i^{\gamma }}
\end{eqnarray}
. Since for the degree $l>k$ only vertices with indices $i>\left\lfloor k^{%
\frac{1}{\alpha }}\left( 1+o\left( k^{\frac{1}{\alpha }}\right) \right)
\right\rfloor $ contribute and in this case the sum $\sum_{l>k}\Pr \left(
d_{in}\left( x\right) =l\mid x\in C_{i}\right) $ is essentially $1,$ we get 
\begin{eqnarray}
F_{k} &=&\sum_{i>\left\lfloor k^{\frac{1}{\alpha }}\left( 1+o\left( k^{\frac{%
1}{\alpha }}\right) \right) \right\rfloor }\sum_{l>k}\Pr \left( d_{in}\left(
x\right) =l\mid x\in C_{i}\right) \frac{c_{1}}{i^{\gamma }} \\
&=&\sum_{i>\left\lfloor k^{\frac{1}{\alpha }}\left( 1+o\left( k^{\frac{1}{%
\alpha }}\right) \right) \right\rfloor }\frac{c_{1}}{i^{\gamma }} \\
&=&\frac{c_{1}}{\left\lfloor k^{\frac{1}{\alpha }}\left( 1+o\left( {}\right)
\right) \right\rfloor ^{\gamma -1}}=\frac{c_{1}}{k^{\frac{\gamma }{\alpha }-%
\frac{1}{\alpha }}\left( 1+o\left( k^{\frac{\gamma }{\alpha }-\frac{1}{%
\alpha }}\right) \right) }
\end{eqnarray}
. Therefore the integrated tail-degree distribution is still of power-law
form.

Next we want to estimate the second moments $M_{2}^{in}$ of the indegree
distribution for values $1<\alpha <\gamma -1$. By definition we have 
\begin{eqnarray}
M_{2}^{in} &=&\sum_{k}\sum_{i}\sum_{x\in C_{i}}k^{2}\Pr \left( d\left(
x\right) =k\mid x\in C_{i}\right) \\
&=&\sum_{k}\sum_{i}k^{2}\Pr \left( d\left( x\right) =k\mid x\in C_{i}\right) 
\frac{c_{1}}{i^{\gamma }}
\end{eqnarray}
and asymptotically since the second moment of a Poisson distribution with
expectation $\lambda $ is given by $\lambda ^{2}-$ $\lambda :$%
\begin{eqnarray}
M_{2}^{in} &\simeq &\sum_{i}\frac{c_{1}}{i^{\gamma }}\cdot \left[ \left( 
\tilde{A}i^{\alpha }\right) ^{2}-\tilde{A}i^{\alpha }\right] \\
&=&\sum_{i}\left( \frac{const}{i^{\gamma -2\alpha }}-\frac{const}{i^{\gamma
-\alpha }}\right)
\end{eqnarray}
. The first term in the sum diverges for $\alpha >\frac{\gamma -1}{2}$ which
is of course also the condition for the total degree second moment to
diverge (in the case $\alpha <1$ the outdegree dominates the indegree and
the second moment diverges for $\gamma <3$). We show in the next paragraph
that this is precisely the condition for absence of an epidemic threshold.

\subsection{The branching process approximation}

In the following we want to analyze the branching process approximation for
the epidemic process on the above described random graph space $\mathcal{G}%
_{n}\left( \alpha ,\gamma \right) $. Due to the absence of local cycles (in
the large $n$ limit) one gets an exact threshold estimation via the
branching process approximation. Let $\mu $ be the transmission probability
that an infection will be transmitted along a given edge of the graphs.
Eliminating edges in $\mathcal{G}_{n}\left( \alpha ,\gamma \right) $ with
probability $1-\mu $ gives a new random graph space $\mathcal{G}_{n}^{\mu
}\left( \alpha ,\gamma \right) $ which we call the epidemic random graph. We
will denote expectations with respect to $\mathcal{G}_{n}^{\mu }\left(
\alpha ,\gamma \right) $ by $\mathbb{E}_{\mu }$ whereas $\mathbb{E}$ is
reserved for expectations with respect to $\mathcal{G}_{n}\left( \alpha
,\gamma \right) $.

Let $T$ $\left( \alpha ,\gamma ,n\right) =\left( a_{ij}\left( \alpha
,n\right) \right) $ be given by 
\begin{equation}
a_{ij}\left( \alpha ,n\right) :=\mathbb{E}_{\mu }\left( \sharp \left\{
y:x\thicksim y\right\} \mid x\in C_{i}\left( n\right) ,y\in C_{j}\left(
n\right) \right)
\end{equation}
\begin{equation}
\text{ \ \ \ \ \ \ \ \ \ \ \ }=\mu \cdot \mathbb{E}\left( \sharp \left\{
y:x\thicksim y\right\} \mid x\in C_{i}\left( n\right) ,y\in C_{j}\left(
n\right) \right)
\end{equation}
and let $a_{ij}\left( \alpha \right) :=\lim_{n\rightarrow \infty
}a_{ij}\left( \alpha ,n\right) .$ Note that $T$ $\left( \alpha ,\gamma
,n\right) $ is the transposed of the branching process transition matrix. In
other words $a_{ij}\left( \alpha \right) $ is the expected number of
infected individuals in group $j$ generated by one infected individual in
group $i$. Actually the exact terms in the asymptotic branching process
matrix are given by $a_{ij}\left( \alpha \right) -1$. Since the difference
by one is irrelevant for our considerations we neglect this term in the
computations. From formula (\ref{55}) we conclude that 
\begin{equation}
a_{ij}\left( \alpha ,n\right) \simeq C_{1}\cdot \mu \cdot \frac{ij^{\alpha
}+ji^{\alpha }}{j^{\gamma }}
\end{equation}
where the constant $C_{1}$ depends only on $\alpha $ and $\gamma $. Absence
of an epidemic threshold for fixed $\alpha $ and $\gamma $ holds exactly if 
\begin{equation}
\lim_{n\rightarrow \infty }\lambda _{\max }\left( T\left( \alpha ,\gamma
,n\right) \right) =\infty \text{ \ \ for }\mu >0
\end{equation}
with $\lambda _{\max }\left( .\right) $ being the largest eigenvalue\ in
modulus.\emph{\ }

The regions in the parameter space for which we will get absence of an
asymptotic epidemic threshold are 1) $2<\gamma \leq 3$\emph{\ }and arbitrary 
\emph{\ }$\alpha $\emph{\ }and 2) the parameters with $\gamma >3$ and $%
\alpha \geq \frac{\gamma -1}{2}$. Finally we get for $\gamma >3$\ and $%
\alpha <\frac{\gamma -1}{2}$\ that there is a size independent threshold.

Since the smallest row sum of a positive matrix gives a lower bound on the
largest eigenvalue it is enough to show that for $\left( b_{lk}\left(
n\right) \right) :=T$ $^{2}\left( \alpha ,\gamma ,n\right) $ and $\alpha $
and $\gamma $ as in case 1) or 2) we have 
\begin{equation}
\min_{l}\sum\limits_{k\geq 1}b_{lk}\left( n\right) \underset{n\rightarrow
\infty }{\rightarrow }\infty .
\end{equation}
Note that $a_{ij}\left( \alpha ,n\right) \simeq a_{ij}\left( \alpha \right) $
and therefore 
\begin{eqnarray}
b_{lk}\left( n\right) &\simeq &\sum\limits_{i}a_{li}\left( \alpha \right)
a_{ik}\left( \alpha \right) \\
&=&\left( \mu C_{1}\right) ^{2}\sum\limits_{i}\frac{li^{\alpha }+il^{\alpha }%
}{i^{\gamma }}\cdot \frac{ik^{\alpha }+ki^{\alpha }}{k^{\gamma }} \\
&=&\left( \mu C_{1}\right) ^{2}(\frac{lk^{\alpha }+l^{\alpha }k}{k^{\gamma }}%
\sum\limits_{i}\frac{i^{\alpha +1}}{i^{\gamma }}+ \\
&&+\frac{lk}{k^{\gamma }}\sum\limits_{i}\frac{i^{2\alpha }}{i^{\gamma }}+%
\frac{l^{\alpha }k^{\alpha }}{k^{\gamma }}\sum\limits_{i}\frac{i^{2}}{%
i^{\gamma }}).
\end{eqnarray}
. From this formula one immediately concludes the absence of an epidemic
threshold in the above described domains since $\sum\limits_{i}\frac{i^{2}}{%
i^{\gamma }}$ respectively $\sum\limits_{i}\frac{i^{2\alpha }}{i^{\gamma }}$
diverge for $\gamma \leq 3$ respectively $\alpha \geq \frac{\gamma -1}{2}$
whereas the divergence of $\sum\limits_{i}\frac{i^{\alpha +1}}{i^{\gamma }}$
is irrelevant since $\frac{\gamma -1}{2}$ $\geq \gamma -2$ for $\gamma \geq
3 $.

To show the existence of an asymptotic threshold for parameters satisfying $%
\gamma >3$\ and $\alpha <\frac{\gamma -1}{2}$ we will demonstrate that for $%
\mu >0$ we have $\lim\limits_{m\rightarrow \infty }\lambda _{\max }\left(
T^{m}\left( \alpha ,\gamma ,n\right) \right) <\infty $ from which the
assertion then follows. Let $\left( g_{lk}^{m}\left( n\right) \right) :=T$ $%
^{m}\left( \alpha ,\gamma ,n\right) $. We will derive a recursion for $%
g_{lk}^{m}\left( n\right) $ as follows. Since 
\begin{equation}
g_{lk}^{m}\left( n\right) \simeq \left( \mu C_{1}\right)
^{m}\sum\limits_{i_{m}...i_{1}}\frac{li_{m}^{\alpha }+i_{m}l^{\alpha }}{%
i_{m}^{\gamma }}\cdot ...\cdot \frac{i_{2}i_{1}^{\alpha }+i_{1}i_{2}^{\alpha
}}{i_{1}^{\gamma }}\cdot \frac{i_{1}k^{\alpha }+ki_{1}^{\alpha }}{k^{\gamma }%
}
\end{equation}
we define variables 
\begin{equation}
g_{lk}^{m}\left( n,x,y\right) :=\left( \mu C_{1}\right)
^{m}\sum\limits_{i_{m}...i_{1}}\frac{li_{m}^{\alpha }+i_{m}l^{\alpha }}{%
i_{m}^{\gamma }}\cdot ...\cdot \frac{i_{2}i_{1}^{\alpha }+i_{1}i_{2}^{\alpha
}}{i_{1}^{\gamma }}\cdot \frac{i_{1}x+yi_{1}^{\alpha }}{k^{\gamma }}
\end{equation}
and observe that 
\begin{equation}
g_{lk}^{m+1}\left( n,x,y\right) =\mu C_{1}\cdot g_{lk}^{m}\left( n,x^{\prime
},y^{\prime }\right)
\end{equation}
where 
\begin{equation}
x^{\prime }=B_{0}x+B_{1}y\text{ \ };\text{ \ }y^{\prime }=B_{2}x+B_{o}y
\end{equation}
and $B_{0}=\sum\limits_{i}\frac{i^{\alpha +1}}{i^{\gamma }}%
,B_{1}=\sum\limits_{i}\frac{i^{2\alpha }}{i^{\gamma }}$ and $%
B_{2}=\sum\limits_{i}\frac{i^{2}}{i^{\gamma }}$. Since $g_{lk}^{m}\left(
n,k^{\alpha },k\right) \simeq g_{lk}^{m}\left( n\right) $ we only have to
show that $\mu C_{1}\cdot \lambda _{\max }\left[ \left( 
\begin{array}{cc}
B_{0} & B_{1} \\ 
B_{2} & B_{o}
\end{array}
\right) \right] <1$ to establish the boundedness of $\lambda _{\max }\left(
T^{m}\left( \alpha ,\gamma ,n\right) \right) $. Clearly if $B_{0},B_{1}$ and 
$B_{2}$ stay bounded as a function of $\ n$ we can match the requirement of
the last sentence. But $B_{0},B_{1}$ and $B_{2}$ stay bounded on the
parameter set $\gamma >3$\ and $\alpha <\frac{\gamma -1}{2}$ from which our
assertion follows.

The result has a nice heuristic explanation. The ''tail''-groups (that is
the groups $C_{i}$ with $i$ very large) act as a hub for the infection.
Indeed given an infected individual in group $C_{k}$ it spreads the
infection into the hub via it's outdegree edges (for $\alpha $ large) from
which the infection is backspread via the indegree edges of the hub to $%
C_{k} $ or any other fixed group. It is precisely this infection path which
is proportional to $\sum\limits_{i}\frac{i^{2\alpha }}{i^{\gamma }}$ and
therefore makes the branching process divergent for $\alpha >2\gamma +1.$

\section{ The very small world inside the small world}

In the following we will show that the distinction between absence and
presence of an epidemic threshold can be exactly related to properties of
the diameter of the giant component $\mathcal{G}_{n}^{0}\left( \alpha
,\gamma \right) $ of $\mathcal{G}_{n}\left( \alpha ,\gamma \right) $. Of
course the diameter of a random graph space is itself a random variable but
it turns out that it has a small variation. For our purpose we therefore
concentrate on the expected diameter. Given a graph $G_{n}$ with $n$
vertices and a subset $A$ of $V_{n}$ we define $diam\left( G_{n},A\right)
:=\max \left\{ d\left( x,y\right) \mid x,y\in A\right\} $ where $d\left(
x,y\right) $ is the distance on the graph $G_{n}$. Let the $\varepsilon -$%
essential diameter of a graph $G_{n}$ be defined as $diam_{\varepsilon
}^{ess}\left( G_{n}\right) :=\min\limits_{A}\left\{ diam\left(
G_{n},A\right) \mid \left| A\right| \geq \left( 1-\varepsilon \right) \left|
V_{n}\right| \right\} $. Similar if we have a sequence of random graph
spaces $\left\{ \mathcal{G}_{n}\right\} $ the quantity $diam_{\varepsilon
}^{ess}\left( \mathcal{G}_{n}\right) $ is now a random variable depending on
the realization $G_{n}\in \mathcal{G}_{n}$.

The striking phenomenon in the case of absence of an epidemic threshold for
the above described parameter region is the following property:

\emph{For all }$\varepsilon \in \left( 0,1\right) $\emph{\ \ there is a }$%
C\left( \varepsilon \right) >0$\emph{\ such that }$\Pr \left\{
diam_{\varepsilon }^{ess}\left( \mathcal{G}_{n}^{0}\left( \alpha ,\gamma
\right) \right) <C\left( \varepsilon \right) \right\} $\emph{\ }$\rightarrow
1$\emph{\ as }$n$ \emph{tends to infinity. }

In other words: A positive fraction of the vertices are located in a cluster
of finite diameter. This cluster has a kind of fuzzy, onion like structure
with no sharp boundary. At the ''center'' of the cluster are the very high
degree groups placed which are entirely contained in the cluster. The union
of shells which are at distance less than $R$ from the center has positive
mass as soon as $R$ gets larger than a critical value $C_{0}$. Increasing $R$
makes this fraction larger and larger but to get the whole giant connected
component in the cluster requires to take $R\rightarrow \infty $ as $n$ goes
to infinity (something between $\log \log n$ and $\log n$ should be the
right growth rate).

Before going into the details let us recall the general philosophy how to
estimate the diameter of a (large) connected component of a random graph
space. Fixing an individual $x$ one tries to estimate the expectation of $%
\Gamma _{k}:=\sharp \left\{ y:d\left( x,y\right) =k\right\} $ as good as
possible. As long as $\sum_{i\leq k}\Gamma _{i}\ll n$ holds and the
variation of the $\Gamma _{k}$ is under control one can use in general a
branching process approximation to get the right order of the expectation of 
$\Gamma _{k}$. The smallest $k_{0}$ such that $\sum_{i\leq k_{0}}\Gamma
_{i}\sim n$ gives then a very good estimation of $\ diam_{\varepsilon
}^{ess} $ (for $\varepsilon $ small).

To see that the above statement is true let us note first that iterating the
branching process matrix $k-$times -denoted in the following by $\left(
t_{ij}^{\left( k\right) }\right) $ and being the transposed of $T^{k}\left(
\alpha ,\gamma \right) -$ gives the expected size of the number of
individuals in group $i$ which are at distance $k$ to a random chosen
individual in group $j$ provided that all expectations $\sum%
\limits_{j}t_{ij}^{\left( k^{\prime }\right) }$ are much smaller than $%
\left| C_{i}\left( n\right) \right| $ for $k^{\prime }\leq k.$ Assume now
that the maximal outdegree respectively the largest group index in our model
scales like $n^{\beta }$ for some 0$<\beta <\frac{1}{\gamma }$ and hence the
size of the group with the maximal outdegree is of order $const\cdot
n^{1-\beta \gamma }$. Since the dominating contribution in the $%
t_{ij}^{\left( k\right) }$ terms comes from $\left( \sum\limits_{l<n^{\beta
}}\frac{const}{l^{\gamma -2\alpha }}\right) ^{k}$ we can safely iterate the
branching process matrix $k-$times as long as $k\beta \left( 1-\gamma
+2\alpha \right) <1-\beta \gamma $ (note that in the ''no-threshold'' case
we always have $1-\gamma +2\alpha >0$)$.$ Let $k_{0}$ be the largest $k$ for
which the above inequality holds. To get a good lower bound on the number of
individuals at distances larger then $k_{0}$ one has to truncate the matrix
properly in such a way that only paths are taken into account which make no
use of groups which are smaller than the entries in the iterated branching
process matrix. Therefore we consider in step $k_{0}+k$ only those groups $%
C_{i}$ for which $i<n^{\beta _{k}}$ with $\beta _{k}:=\frac{1}{\left(
1-\gamma +2\alpha \right) \left( k_{0}+k\right) +\gamma }$. The leading term
after $k_{0}+k$ iterations is now given by $n^{k_{0}\beta \left( 1-\gamma
+2\alpha \right) }\prod\limits_{l\leq k}n^{\beta _{l}\left( 1-\gamma
+2\alpha \right) }=n^{k_{0}\beta \left( 1-\gamma +2\alpha \right)
+\sum_{l\leq k}\frac{\left( 1-\gamma +2\alpha \right) }{\left( 1-\gamma
+2\alpha \right) \left( k_{0}+k\right) +\gamma }}$. Since the sum in the
exponent diverges there is a size independent $k_{1}$ such that $%
n^{k_{0}\beta \left( 1-\gamma +2\alpha \right) }\prod\limits_{l\leq
k_{1}}n^{\beta _{l}\left( 1-\gamma +2\alpha \right) }\simeq n$ and
consequently a finite fraction of the vertex set stays within a bounded
distance.

Let us finally demonstrate that there cannot be a $C^{\ast }$ such that
essentially all vertices of the giant component are within a distance less
than $C^{\ast }.$ Fix $f>0$ and an individual $x$ with $d\left( x\right) <f.$
From formula (2.4.) it follows that there is a $n$ independent $p\left(
f\right) >0$ such that the probability that $x$ has only partners $y$ with $%
d\left( y\right) <f$ \ is larger $p\left( f\right) >0.$ Therefore the
probability that at distance less equal $C^{\ast }$ from $x$ there are only
individuals with degree less than $f$ is larger than $p\left( f\right)
^{f^{C^{\ast }+1}}.$ Since there are $const\left( f\right) \cdot n$
individuals with degree less than $f$ there is a small but positive fraction
of individuals which are at distance larger than $C^{\ast }$ from the high
degree groups for arbitrary large $C^{\ast }.$ In a similar way one can show
that there are individuals with distance of order $\ln \ln n$ (but those
form only a set of asymptotic zero density).

\section{A toy example}

In the following section we want to discuss a model in which the
degree-distribution is somehow extreme, namely there are only two types of
individuals with sizes $C_{2}$ and $C_{f\left( n\right) }$. The index $2$
and $f\left( n\right) $ stands for the degree of the two individual-types
the last one depends on the size $n$ of the vertex set. We will make the
general requirement that the second moment of the degree distribution
diverges whereas the first moment converges: 
\begin{equation}
\lim \sup f\left( n\right) \frac{\left| C_{f\left( n\right) }\right| }{n}%
<\infty  \label{66}
\end{equation}
\begin{equation}
\left[ f\left( n\right) \right] ^{2}\frac{\left| C_{f\left( n\right)
}\right| }{n}\rightarrow \infty .
\end{equation}

We will look at the epidemic process initiated by one randomly infected
individual in group $C_{f\left( n\right) }$ and ask wether one has a
persistent epidemics for arbitrary small transmission probability. For this
model the question of existence of an epidemic threshold is only interesting
in the above conditional sense since due to our assumptions the asymptotic
fraction of $C_{f\left( n\right) }$ individuals on the total population
tends to zero and no epidemics can be persistent among the $C_{2}$
population for sufficiently small transmission probabilities.

We will investigate two different structures of edge formations.

\subsection{Case A}

In this subsection we want to separate the elements of $C_{f\left( n\right)
} $ as far as possible from each other which gives the epidemiologically
''safest'' structure in the sense that unless the first moments of the
degree-distribution ''almost'' diverge one would expect an undercritical
situation if the transmission probability is sufficiently small (note that
if the first moment of the degree-distribution diverges one gets trivially
no threshold). We want to separate all elements of $C_{f\left( n\right) }$
by chains of uniform length $l\left( n\right) $ consisting of elements of $%
C_{2}$. Since we have $f\left( n\right) \frac{\left| C_{f\left( n\right)
}\right| }{2}$ chains (we drop all effects due to integer part requirements
since they are asymptotically irrelevant) we get for the length $l\left(
n\right) \cong \frac{2\left| C_{2}\right| }{\left| C_{f\left( n\right)
}\right| f\left( n\right) }+1.$ The probability to transmit an infection
along such a chain is $\mu ^{l\left( n\right) }$ and hence the expected
number of individuals in $C_{f\left( n\right) }$ which will be infected from
one initially infected individual from group $C_{f\left( n\right) }$ equals $%
f\left( n\right) \mu ^{l\left( n\right) }$. To get absence of a threshold we
need therefore $f\left( n\right) \mu ^{l\left( n\right) }>1$ asymptotically.
This is equivalent to $\log f\left( n\right) -c\frac{\left( n-\left|
C_{f\left( n\right) }\right| \right) }{\left| C_{f\left( n\right) }\right|
f\left( n\right) }>0$ with $c=2\log \frac{1}{\mu }$ from which by (\ref{66}) 
$O\left( n\right) \log f\left( n\right) >c\cdot \left( n-\left| C_{f\left(
n\right) }\right| \right) $ follows. First we observe that it is impossible
to find exponents $\beta ,\delta \in \left( 0,1\right) $ s.t. $f\left(
n\right) :=n^{\beta }$ and $\left| C_{f\left( n\right) }\right| :=n^{\delta
} $ and $f\left( n\right) \mu ^{l\left( n\right) }\rightarrow \infty $ under
the assumption $\beta +\alpha <1$. Note that $\beta +\alpha =1$ is just the
borderline where the expectation of the degree starts to become large.

As a conclusion from the considerations in this section one should probably
take, that it is the balance between separation length and degree growth
which matters for the epidemiological threshold-question. So if the elements
from $C_{2}$ would not be as uniformly spread we should look at the
following quantities: $\varphi _{l}\left[ f\left( n\right) \right] :=\mathbb{%
E}_{x}\left[ \sharp \left( y\in C_{f\left( n\right) }:dist_{0}\left(
x,y\right) =l\right) \right] $ where $dist_{0}$ denotes the length of the
shortest path between $x\in C_{f\left( n\right) }$ and $y$ which entirely
consists of individuals from $C_{2}.$ For getting an overcritical epidemics
we therefore need $A\left( n\right) :=\sum\limits_{l\geq 0}\varphi _{l}\left[
f\left( n\right) \right] \mu ^{l}>1.$

\subsection{Case B}

Assume now that we have stationary asymptotic probabilities $p_{00},$ $%
p_{01},$ $p_{10}$ and $p_{11}$ for the events that a single random chosen
edge from an $x\in C_{f\left( n\right) }$ points to an element from $%
C_{f\left( n\right) }$ ($p_{00}$) or to an element from $C_{2}$ ($p_{01}$)
respectively that an edge from an element from $C_{2}$ points to $C_{f\left(
n\right) }$ ($p_{10}$) or to an element from $C_{2}$ ($p_{11}$). There are
three conditions on the choice of the $p_{ij}$ namely $%
p_{00}+p_{01}=1,p_{10}+p_{11}=1$ and $p_{01}\cdot f\left( n\right) \left|
C_{f\left( n\right) }\right| =p_{10}\cdot 2\left| C_{2}\right| $. Since the
first moments are bounded we have $\left| C_{f\left( n\right) }\right|
=o\left( n\right) $ and therefore $\left| C_{2}\right| \sim n$. To get the
threshold condition we have to estimate the expected number of secondary
infections caused by an infected individual from $C_{f\left( n\right) }$
which is precisely given by the quantity $A\left( n\right) $. To compute the
numbers $\varphi _{l}\left[ f\left( n\right) \right] $ observe that 
\begin{equation}
\Pr \left( \text{a fixed chain starting at }x\text{ has length }l,l>1\right)
=p_{01}p_{11}^{l-2}p_{10}
\end{equation}
and $\Pr \left( \text{a fixed chain starting at }x\text{ has length }%
1\right) =p_{00}$. Therefore we have 
\begin{eqnarray}
\sum_{l\geq 1}\varphi _{l}\left[ f\left( n\right) \right] \cdot \mu ^{l}
&=&f\left( n\right) \left( \mu p_{00}+\sum_{l>1}\mu
^{l}p_{01}p_{11}^{l-2}p_{10}\right) \\
&=&f\left( n\right) \mu p_{00}+f\left( n\right) \mu ^{2}p_{01}p_{10}\frac{1}{%
1-\mu p_{11}}
\end{eqnarray}

(we dropped the $n$-dependence of the $p_{ij}$). Clearly the direct coupling
between the $C_{f\left( n\right) }$ individuals has to be of order at most $%
\frac{1}{f\left( n\right) }$ to avoid a nontrivial epidemics inside the $%
C_{f\left( n\right) }$ population which implies $f\left( n\right) \left|
C_{f\left( n\right) }\right| \ll C_{2}$ and in turn $p_{10}\rightarrow 0$.
Using the relations between the $p_{ij}$ we have $A>\mu ^{2}\cdot \frac{%
f^{2}\left( n\right) \left| C_{f\left( n\right) }\right| }{2\left|
C_{2}\right| }$. From the divergence of the second moments one gets $\frac{%
f^{2}\left( n\right) \left| C_{f\left( n\right) }\right| }{2\left|
C_{2}\right| }\rightarrow \infty $ and hence absence of an epidemic
threshold in this probabilistic setting. It remains to estimate the
asymptotic number $I\left( n\right) $ of infected individuals which equals
the size of the largest component of the epidemic random graph. We first
estimate the number of infected individuals in $C_{f\left( n\right) }$
denoted by $I_{f}$ via the stationarity equation $\left| C_{f\left( n\right)
}\right| \left( 1-\frac{1}{A\left( n\right) }\right) =I_{f}\left( n\right) $%
. Furthermore an infected individual from $C_{f\left( n\right) }$ infects in
mean $\sum_{l\geq 1}f\left( n\right) \cdot \mu ^{l}=f\left( n\right) \frac{1%
}{1-\mu }$ individuals from $C_{2}$ from which one obtains the following
asymptotic bounds on the number of infected individuals: $\frac{1}{2}\left|
C_{f\left( n\right) }\right| \left( 1-\frac{1}{A\left( n\right) }\right)
f\left( n\right) \frac{1}{1-\mu }\leq I\left( n\right) \leq \left|
C_{f\left( n\right) }\right| \left( 1-\frac{1}{A\left( n\right) }\right)
f\left( n\right) \frac{1}{1-\mu }$. Therefore unless one has $%
\lim_{n\rightarrow \infty }\frac{\left| C_{f\left( n\right) }\right| f\left(
n\right) }{\left| C_{2}\right| }>0$ (in this case one gets trivially absence
of a threshold since $p_{00}>0$) the size of the asymptotic fraction of
infected individuals in the case of absence of a threshold still converges
to zero.

\section{Conclusions and comments}

We have shown that for the preferential choice model described in this paper
the divergence of the second moment is precisely the condition for absence
of an epidemic threshold even in the case when the asymptotic distribution
is not of a power-law form (but has gaps). In terms of graph geometry this
is related to the presence or absence of a massive finite diameter cluster
which carries the epidemics. It seems that for the case of an exponent
larger than \ three one always has a size independent threshold if one
additionally requires that the size of the largest component of the epidemic
random graph is proportional to $\mu \cdot n$. Note that it is easy to get
absence of threshold for exponents larger than three if one introduces
horizontal preferences in the partner choice (for instance a strong bias to
make connections into the own group) but in this case the size of the
largest component scales like $\mu ^{1+\epsilon }\cdot n$ with an $\epsilon
>0$. It is a much more difficult task to estimate exactly the size of the
giant component on the epidemic random graph which will be the subject of a
more mathematical forthcoming paper \cite{11}. Certainly it scales like $%
const\cdot \mu \cdot n$. It would be furthermore very interesting to have
random network growth principles which produce in a natural way the scale
free graphs we described in this paper. For some results in this direction
see \cite{12}. In any case the model seems to be interesting in describing
computer networks and associated spread of viruses on it. Caution is
required if one wants to draw conclusions for sexually transmitted diseases
like HIV on social networks even when these networks have a power law tail
distribution for the degree. The main point here is that the transmission
probability scales with the degree for small transmission probabilities-
like in the AIDS-epidemics- since high degree individuals \ spend
necessarily less time with each single partner \cite{6}.

Finally we would like to mention that there is a natural variant of our
model by defining independent directed edge probabilities between $x\in
C_{i} $ and $y\in C_{j}$ as $P_{ij}=A_{\alpha }\frac{j^{\alpha }i}{n}$ with
the same $A_{\alpha }$ as above. For this model one gets the same threshold
statements since it has the same branching process transition matrix. For
some alternative strategies to choose partners and the corresponding
threshold estimations see also\cite{21}\cite{31}.

\end{document}